\documentclass[a4paper]{jpconf}
\usepackage{graphicx}
\usepackage{epstopdf}
\usepackage{color}

\begin{document}
\title{The acoustic low-degree modes of the Sun measured with 14 years of continuous GOLF \& VIRGO measurements}
\author{R A Garc\'\i a$^1$, D. Salabert$^{2,3}$, J. Ballot$^4$, K. Sato$^1$, S. Mathur$^5$, A. Jim\'enez$^{2,3}$
}
\address{$^1$ Laboratoire AIM, CEA/DSM-CNRS-Universit\'e Paris Diderot; CEA, IRFU, SAp, F-91191, Gif-sur-Yvette, France}
\address{$^2$ Instituto de Astrof\'isica de Canarias, E-38200 La Laguna, Tenerife, Spain}
\address{$^3$ Dept. de Astrof\'isica, Universidad de La Laguna, E-38206 La Laguna, Tenerife, Spain}
\address{$^4$ Laboratoire d'Astrophysique Toulouse-Tarbes - OMP, 14 avenue Edouard Belin, 31400 Toulouse, France}
\address{$^5$ High Altitude Observatory, 3080 Center Green Drive, Boulder, CO, 80302, USA}

\ead{rgarcia@cea.fr, salabert@iac.es, jballot@ast.obs-mip.fr, kumiko.sato@cea.fr, savita@ucar.edu, ajm@iac.es}




\begin{abstract}	
The helioseismic Global Oscillation at Low Frequency (GOLF) and the Variability of solar Irradiance and Gravity Oscillations (VIRGO) instruments onboard SoHO, have been observing the Sun continuously for the last 14 years. In this preliminary work, we characterize the acoustic modes over the entire p-mode range in both, Doppler velocity and luminosity, with a special care for the low-frequency modes taking advantage of the stability and the high duty cycle of space observations. 
\end{abstract}

\section{Introduction}
Helio- and astero-seismology are the only tools available to pierce inside the stars and provide stratified information of the structure and dynamics of their, otherwise, hidden interiors. This information is embedded in the resonant modes stochastically excited --in the case of solar-like stars-- by turbulent motions taking place in the convective layers. In particular, the acoustic (p) modes with the largest horizontal scales --lowest angular degree $\ell$-- are the only ones that can be measured in Sun-as-a star instruments such as GOLF [1] and VIRGO [2] onboard the Solar and Heliospheric Observatory (SoHO) spacecraft. These global modes are very important since they can propagate through the deepest layers of the stars and penetrate inside their cores, giving for example information on the internal structure [e.g. 3], the internal rotation [e.g. 4], and the magnetic activity cycles, either in the Sun [e.g. 5] or in other stars [6]. In this work, we analyze observations collected with these two instruments to characterize the p-mode properties.

\section{Data analysis and results}
We analyzed nearly 14 years of data collected by the GOLF and VIRGO instruments: 5163 days of GOLF velocity time series [7,8] from April 11, 1996 to May 30, 2010 with a duty cycle, dc=95.4~\%; and 5154 days of intensity data from the three VIRGO Sun photometers (SPM) at 402, 500, and 862~nm from April 11, 1996 to May 21, 210 (dc = 95.2~\%). GOLF is a resonant scattering spectrophotometer that measures the line-of-sight integrated Doppler velocity of the Sun. It has been designed to be very stable at low frequency, allowing, for example, to indirectly detect gravity modes in the Sun [9,10]. On the other hand, the overall noise level of VIRGO/SPM is higher than in GOLF: the signal-to-noise ratio (SNR) being smaller, the frequency range on which we can extract reliable estimates of p-mode characteristics is reduced.

\begin{figure}
\includegraphics[width = 0.5\textwidth]{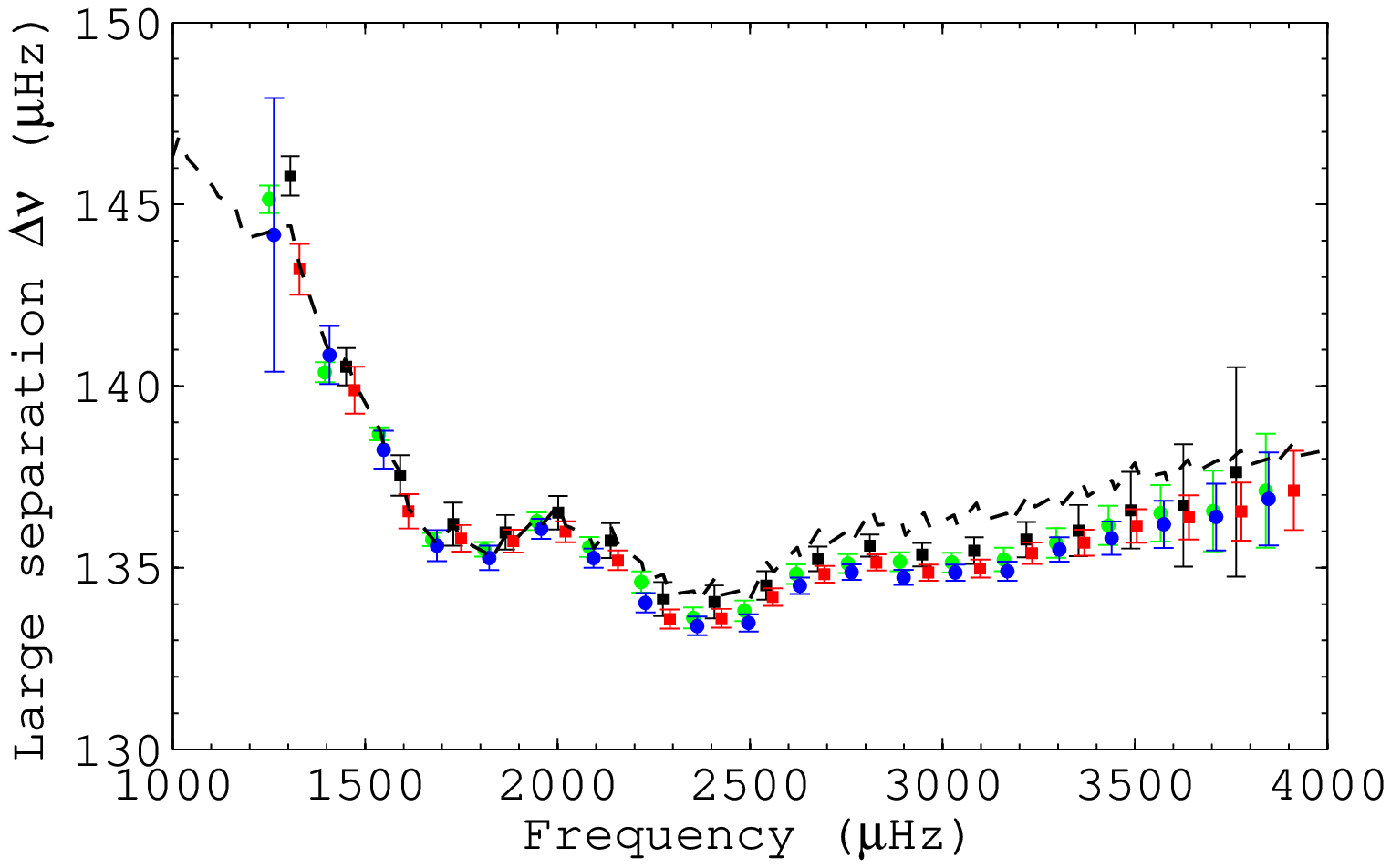}	
\includegraphics[width = 0.5\textwidth]{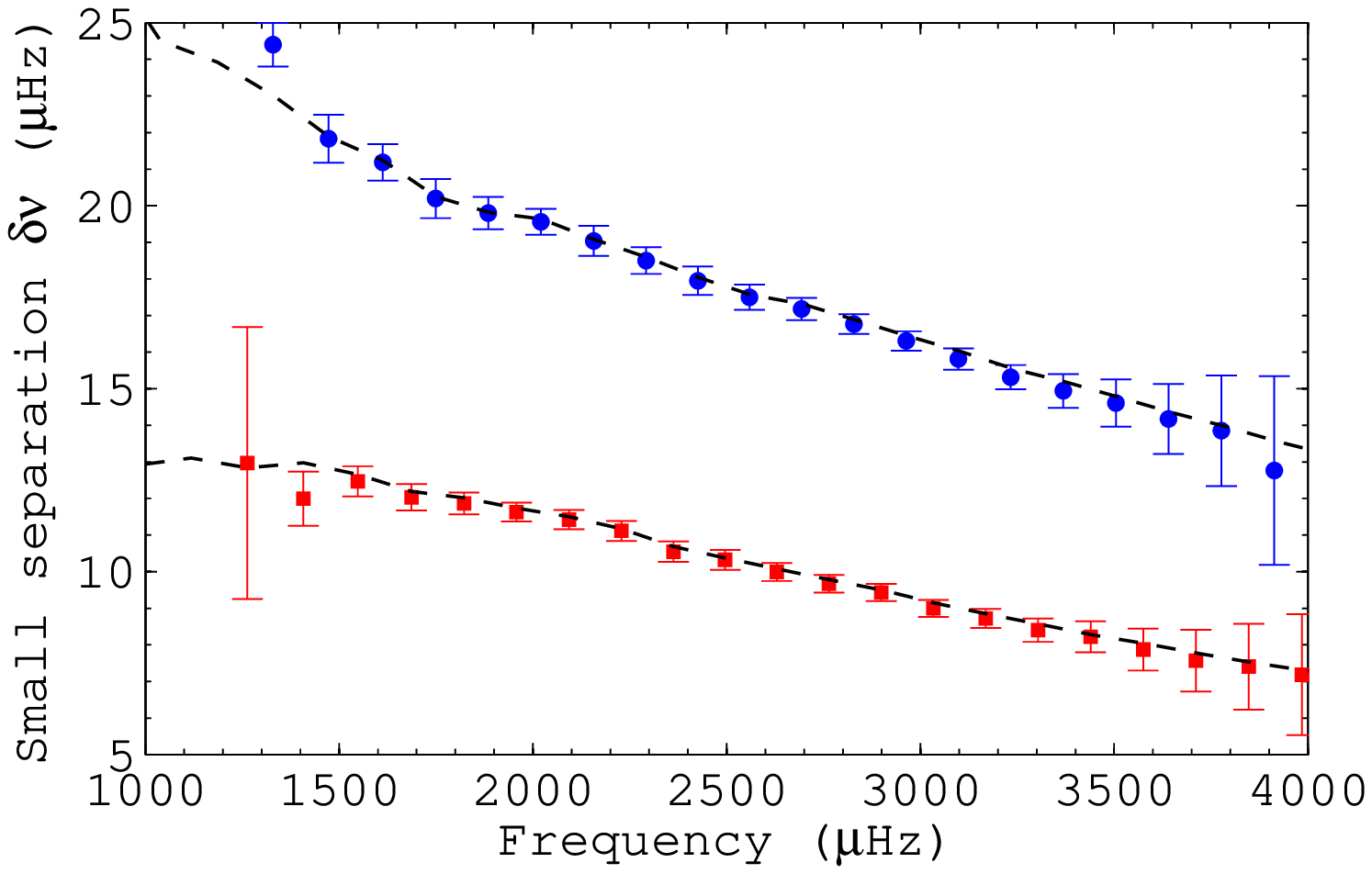}
\includegraphics[width = 0.5\textwidth]{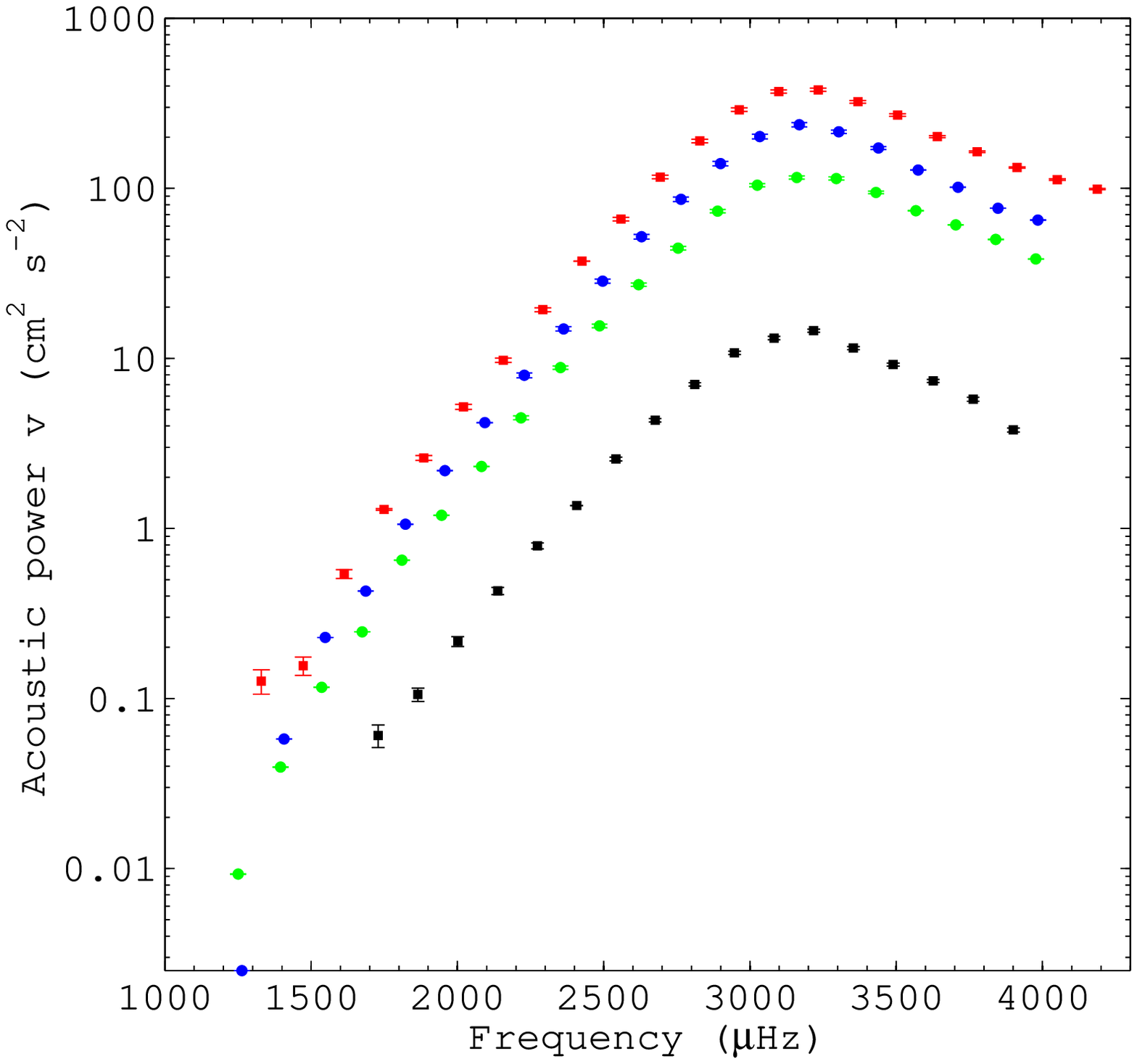}	
\includegraphics[width = 0.5\textwidth]{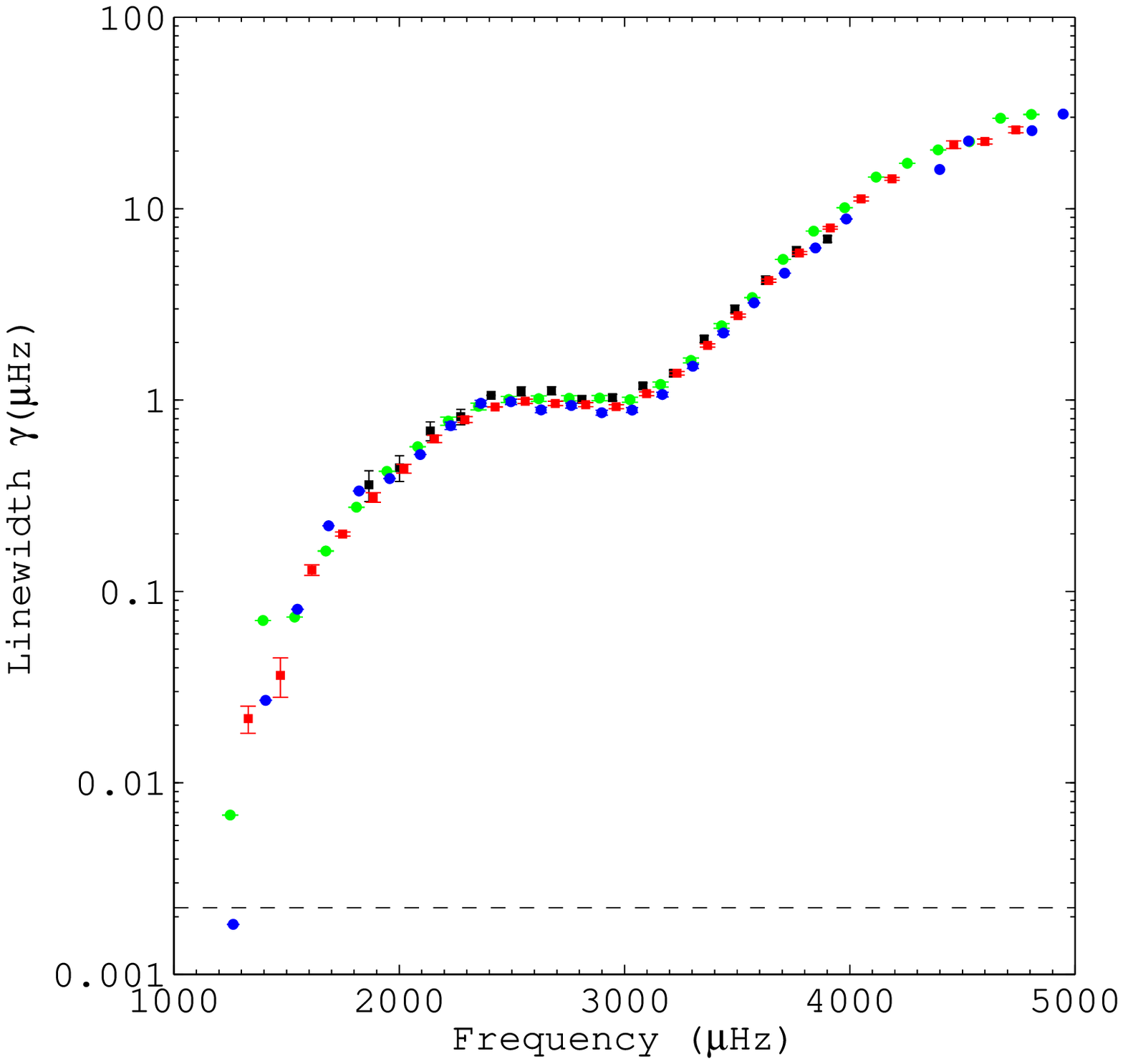}
\includegraphics[width = 0.5\textwidth]{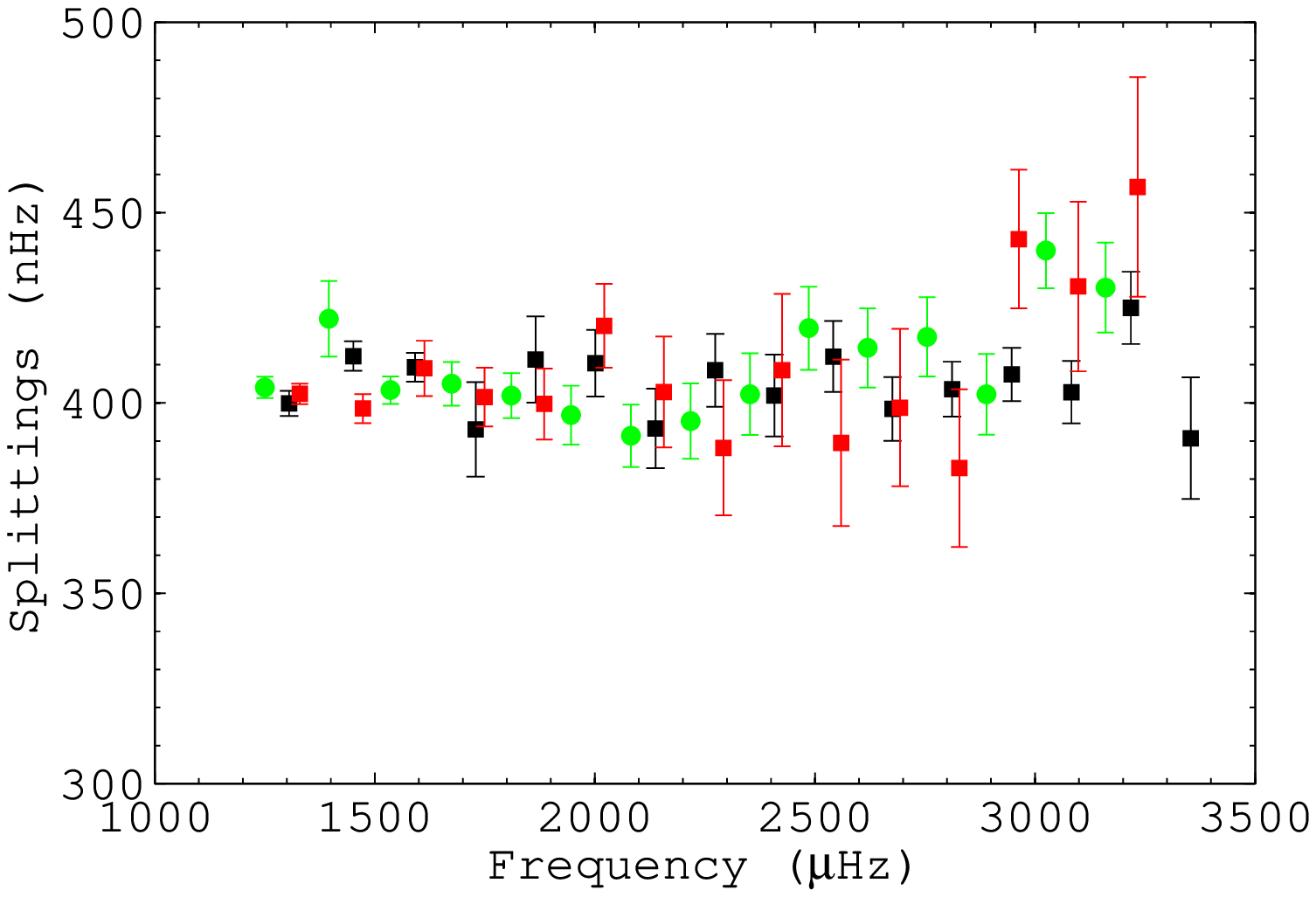}	
\includegraphics[width = 0.5\textwidth]{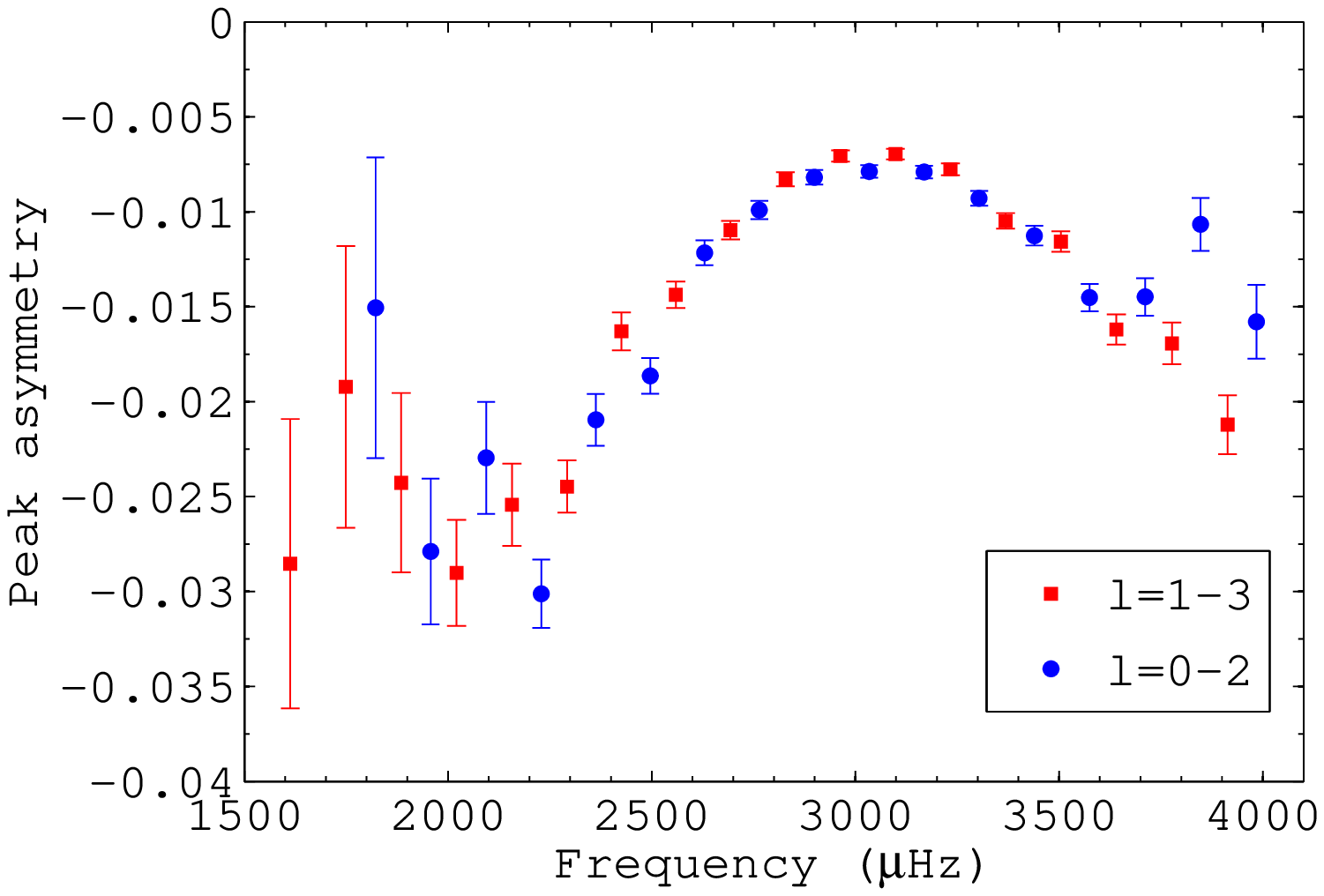}
\caption{Top left: large separation as a function of frequency calculated from the fitted GOLF frequencies (same legend than in VIRGO/SPM plots). The formal errors were multiplied by a factor 10. The dashed lines correspond to the theoretical values using the Saclay seismic model [18]. Top right: small separation ($l=0-2$ modes in red, $l=1-3$ modes in blue). The formal errors were multiplied by a factor 10. Middle left: Full amplitudes (in units of ppm$^2$). Middle right: Linewidths. The horizontal dashed line corresponds to the frequency resolution. Bottom left: splittings. Above 3500 $\mu$Hz, the splittings were fixed to 400 nHz. Bottom right: Asymmetry. The increase below 2000 $\mu$Hz could not be real due to the reduction in the SNR and the fewer number of points defining the profile.
\label{fig:GOLF}}
\end{figure}

\begin{figure}
\includegraphics[width = 0.5\textwidth]{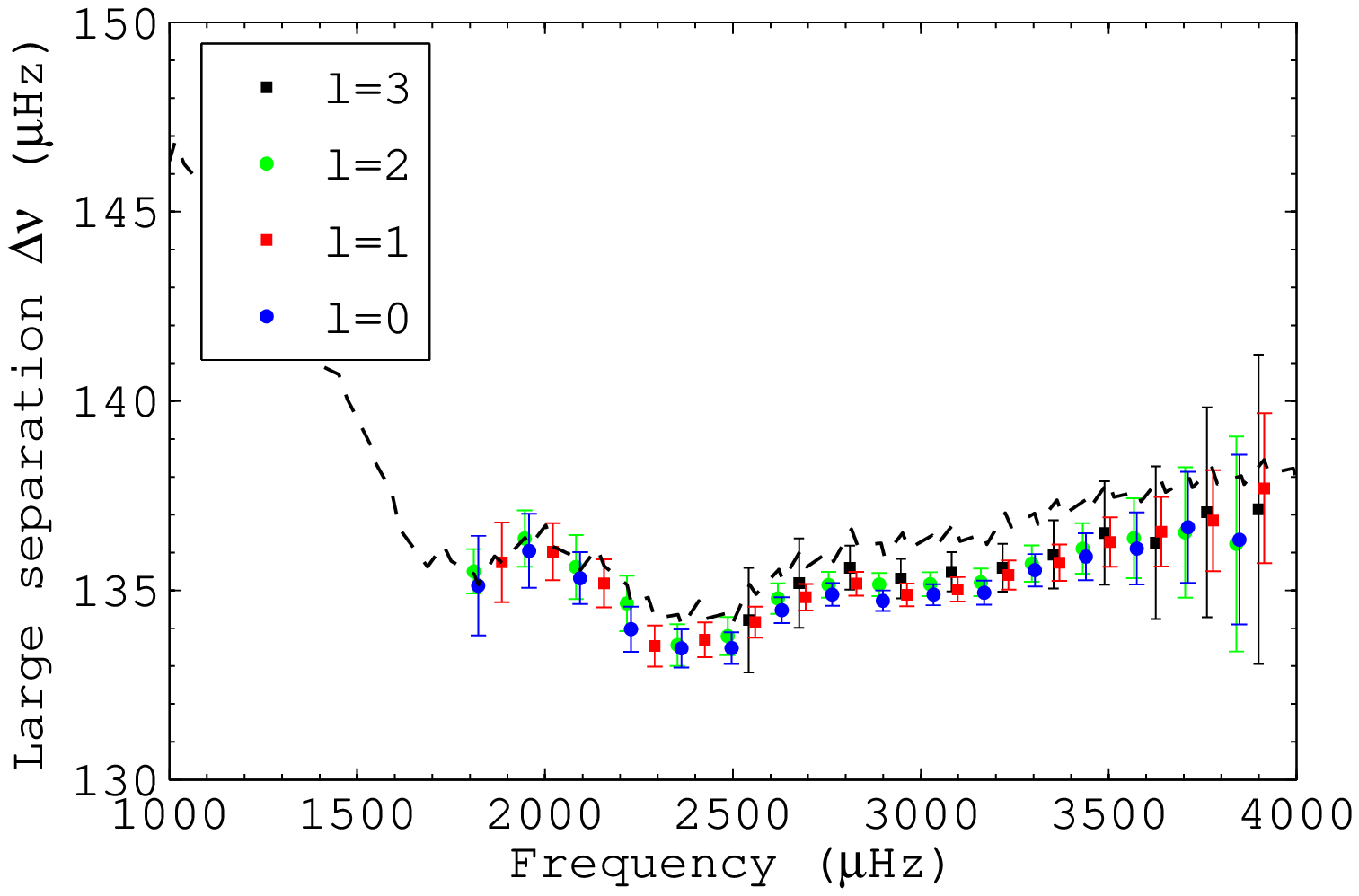}	
\includegraphics[width = 0.5\textwidth]{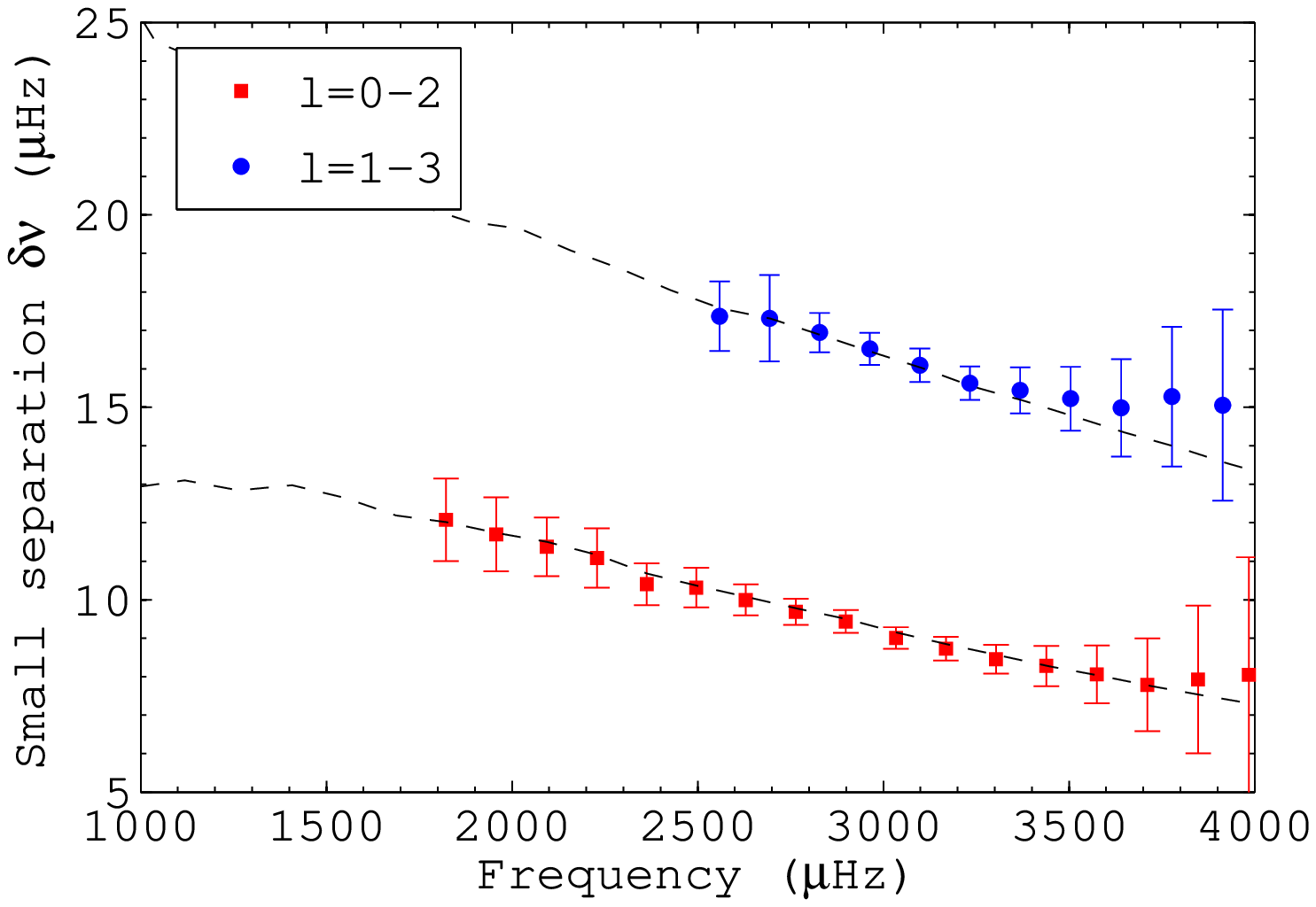}
\includegraphics[width = 0.5\textwidth]{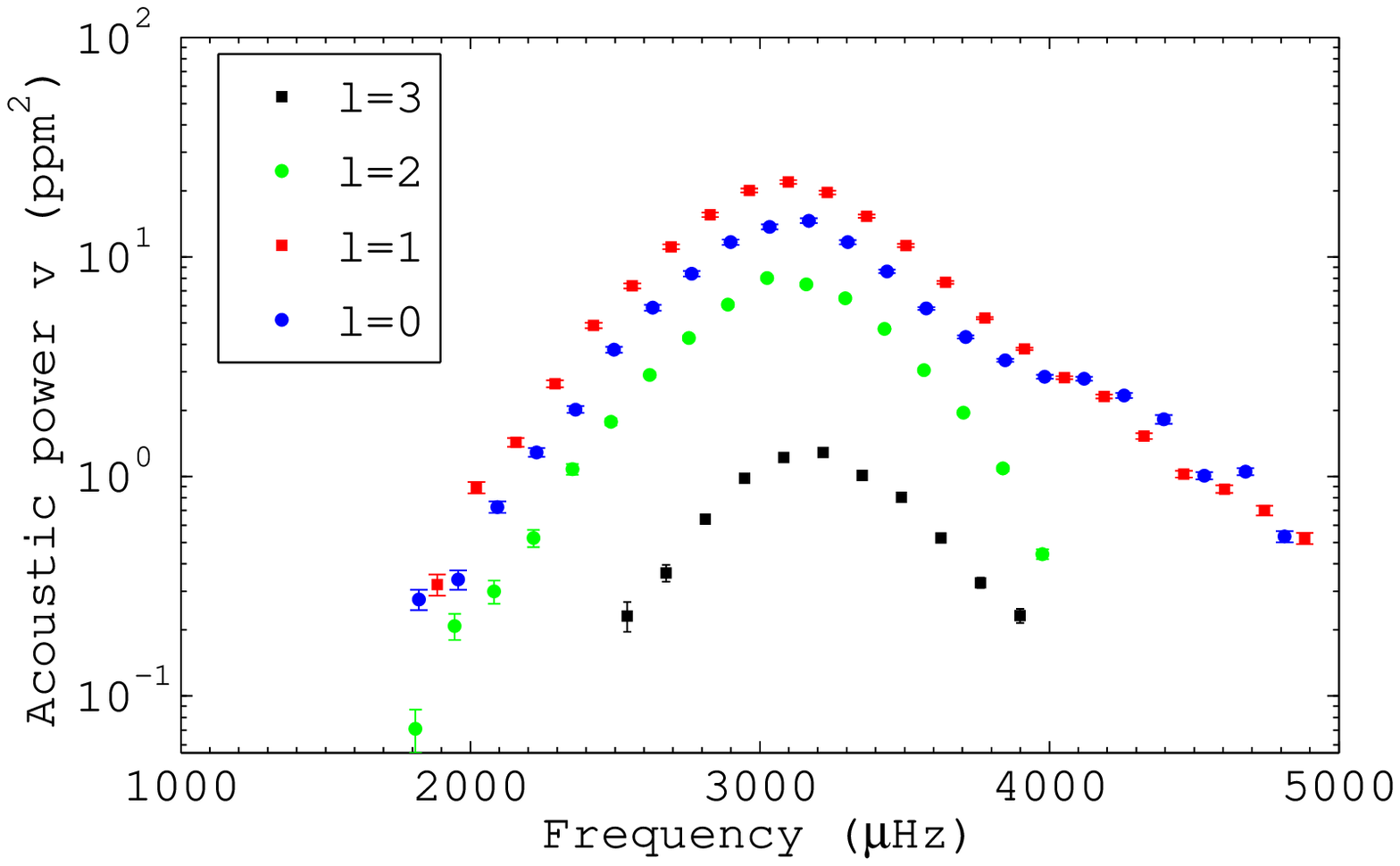}	
\includegraphics[width = 0.5\textwidth]{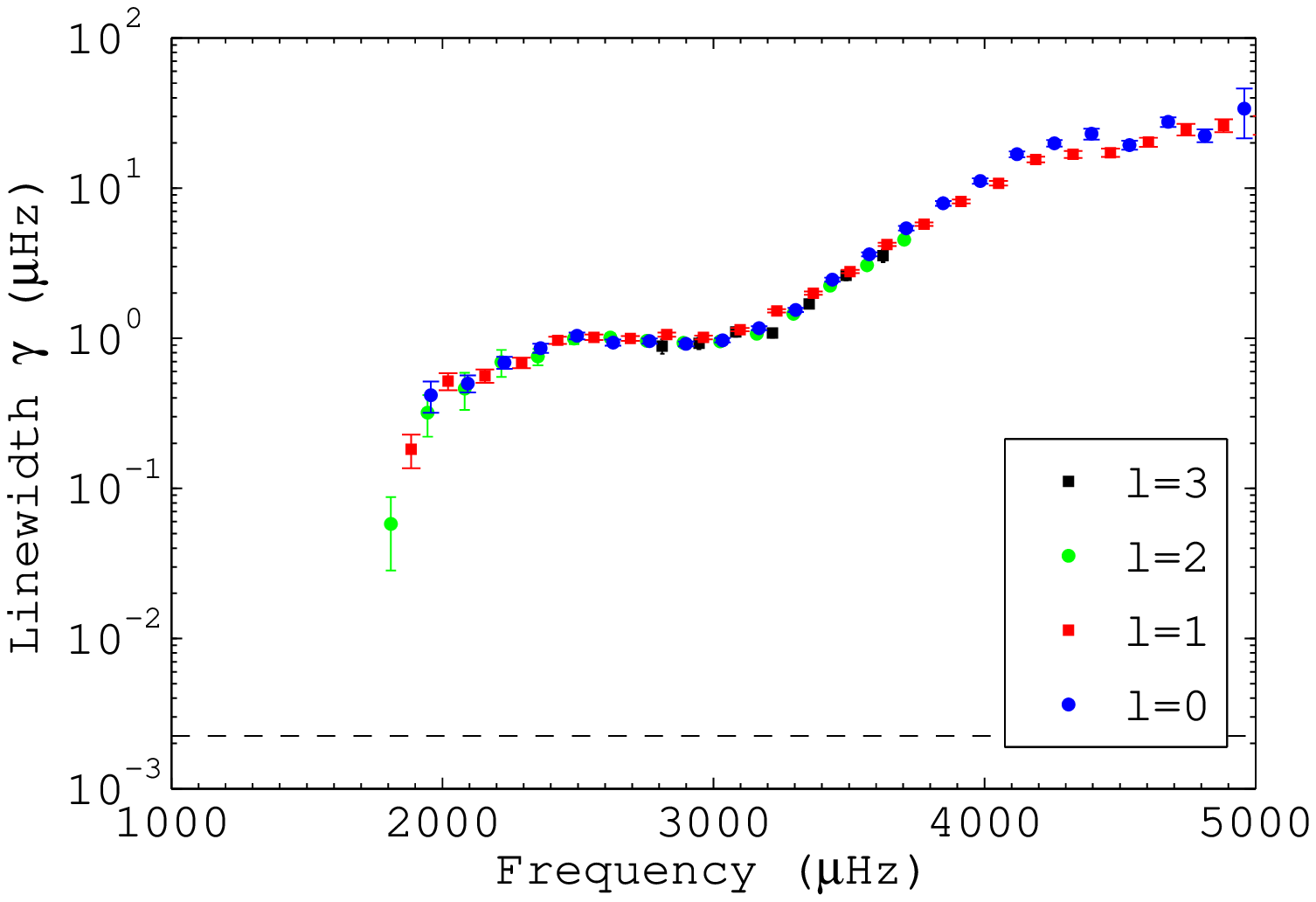}
\includegraphics[width = 0.5\textwidth]{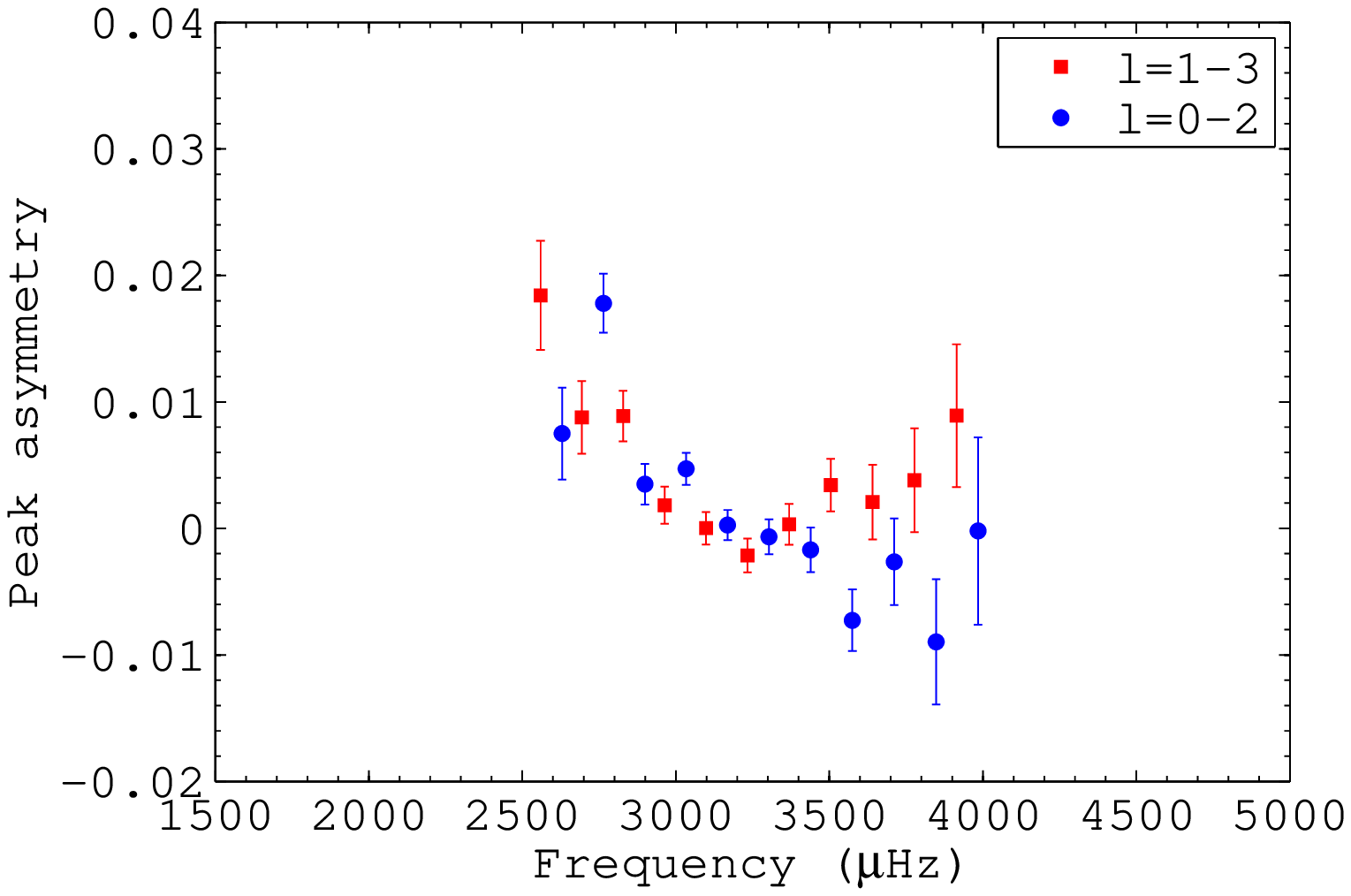}
\includegraphics[angle=90,width = 0.47\textwidth]{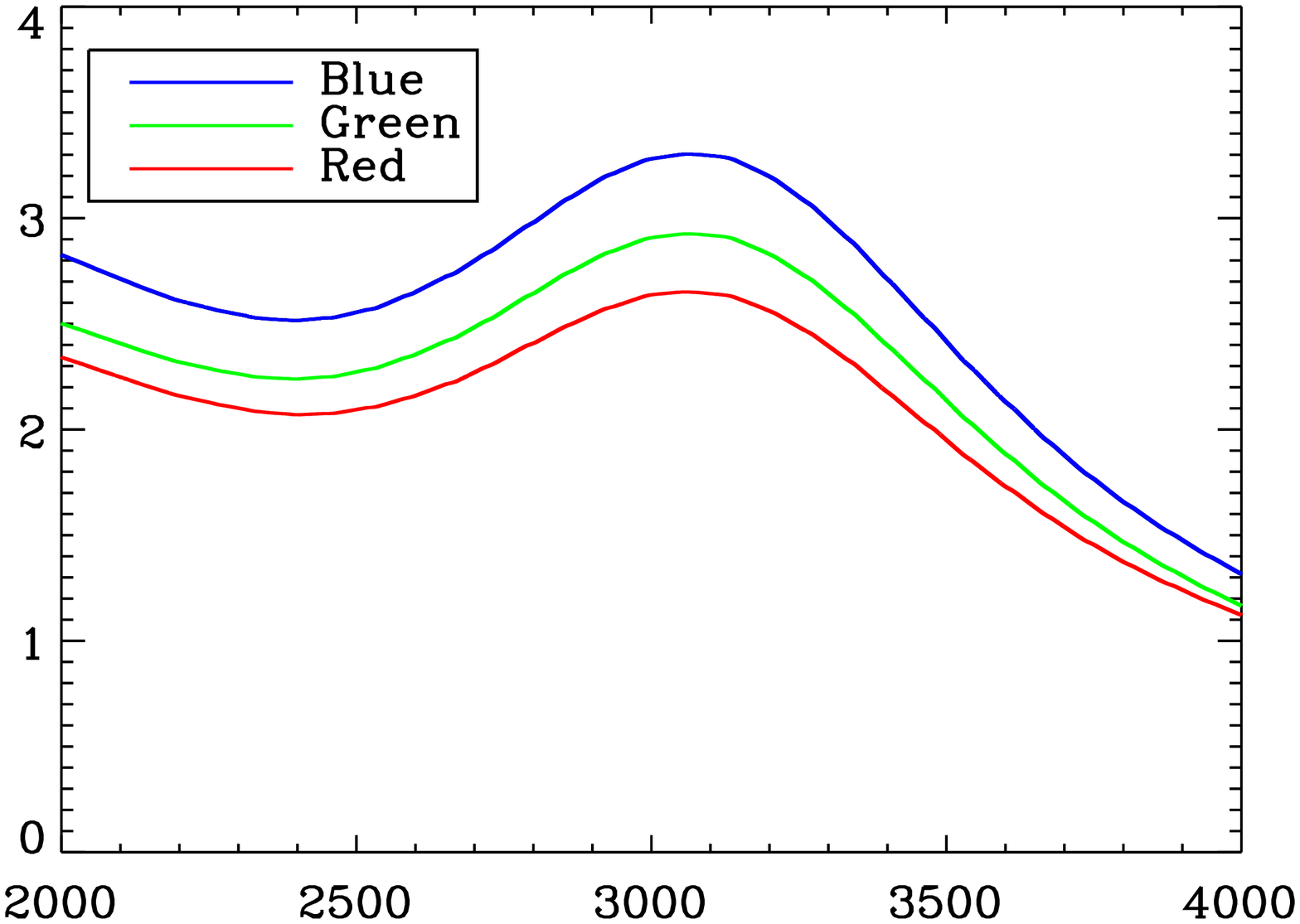}
\caption{Top left:  large separation as a function of frequency calculated from the fitted VIRGO/SPM frequencies. Top right: small separation. Middle left: Full amplitudes (in units of ppm$^2$). Middle right: Linewidths. Bottom left: Asymmetry. Bottom right: average maximum rms amplitudes per radial mode for the three VIRGO/SPM channels. Due to the small $l = 3$ visibility in the VIRGO/SPM data, these modes do not perturb the  $l = 1$ and the linewidth and acoustic power of the $l=1$ modes could be fitted up to 5000 $\mu$Hz. 
\label{fig:Virgo}}
\end{figure}

To characterize the p modes, we computed the power spectrum density (PSD) of the entire time series in order to maximize the frequency resolution ($\sim$ 2.24 nHz). Therefore, the obtained linewidths of the modes could be slightly overestimated because of the shift of the modes during the solar activity cycle [e.g. 5]. The acoustic modes were described using asymmetric Lorentzian profiles, extracted using a classical maximum likelihood method. The fits were performed by a multi-step iterative method [11,12].  Figures~\ref{fig:GOLF} and \ref{fig:Virgo} show the p-mode parameters as a function of frequency extracted from GOLF and VIRGO/SPM respectively between 1000 to 4000 $\mu$Hz, and up to 5000 $\mu$Hz when possible (the splitting is left constant (400 nHz) for modes above 3.5 mHz). For instance, preliminary results of the linewidths extracted from GOLF are given up to 5000 $\mu$Hz. Also, due to the smaller $l = 3$ visibility in the VIRGO/SPM data, the linewidth and acoustic power of the $l = 1$ mode could be fitted up to 5000 $\mu$Hz, the mode blending at high frequency being the limiting factor in radial velocity measurements. Reliable estimates below 1800 $\mu$Hz could not be properly extracted in VIRGO/SPM data because the smaller SNR at low frequency in intensity measurements compared to radial velocity measurements.

We also computed the average maximum amplitude per radial mode of the Sun (Fig~\ref{fig:Virgo} bottom right) for the three VIRGO/SPM channels, as it is commonly done in asteroseismology [e.g. 13,14].  
The maximum amplitudes were corrected by the instrumental response function using the values given by [15] for the different channels.

Some new developments will be addressed in the near future: for instance, we plan to improve the analysis for the low-SNR, low-frequency modes by using specific techniques, i.e., the collapsograms [16]. We will also provide the oscillation parameters corrected from the solar activity by fitting simultaneously the parameters and their variation with activity [17].

\ack
SoHO is a space mission of international cooperation between ESA and NASA. DS acknowledges the support the Spanish National Research Plan grant PNAyA 2007-62650, J.B. the ANR SIROCO, and R.A.G the CNES grant. NCAR is supported by the National~Science~Foundation.

\end{document}